\documentclass[twocolumn]{svjour3}          
\smartqed  
\usepackage{graphicx,natbib,aas_macros}
\def\pdot {\dot \textrm{P}}

\def\ltsima{$\; \buildrel < \over \sim \;$}
\def\lsim{\lower.5ex\hbox{\ltsima}}
\def\gtsima{$\; \buildrel > \over \sim \;$}
\def\gsim{\lower.5ex\hbox{\gtsima}}
\def\msun{~M_{\odot}}

\def\smc {CXO\,J0100$-$7211}
\def\uu     {4U\,0142$+$61}
\def\zeroq {SGR\,0418$+$5729}
\def\zeroc {SGR\,0501$+$4516}
\def\lmc   {SGR\,0526$-$66}
\def\oo     {1E\,1048.1$-$5937}
\def\qui    {1E\,1547.0$-$5408}
\def\psr  {PSR\,J1622$-$4950}
\def\sedici{SGR\,1627$-$41}
\def\cxo {CXO\,J1647$-$4552}
\def\rxs {1RXS\,J1708$-$4009}
\def\ctb {CXO\,J1714$-$3810}
\def\zerosei  {SGR\,1806$-$20}
\def\xte     {XTE\,J1810$-$197}
\def\swia  {Swift\,J1822$-$1606}
\def\diciotto {SGR\,1833$-$0832}
\def\swib  {Swift\,J1834$-$0846}
\def\kes       {1E\,1841$-$045}
\def\zerozero {SGR\,1900$+$14}
\def\ee        {1E\,2259$+$586}

\def\xdzerosette {RX~J0720.4$-$3125}

\begin{document}

\title{Pulsars and Magnetars
}


\author{Sandro Mereghetti
}


\institute{S. Mereghetti \at
              INAF, IASF-Milano \\
              v. E.Bassini 15, I-20133 Milano\\
              Italy\\
              \email{sandro@iasf-milano.inaf.it}           
}


\date{ Proceedings of the 26th Texas Symposium on Relativistic Astrophysics, Sao Paulo, December 16-20 2012}

\maketitle

\begin{abstract}
The high-energy sources known as anomalous X-ray pulsars (AXPs) and soft $\gamma$-ray repeaters (SGRs)
are well explained as magnetars:  isolated neutron stars powered by their own magnetic energy.
After explaining why it is generally believed that the traditional energy sources at work in other neutron stars
(accretion, rotation, residual heat) cannot power the emission of  AXPs\slash SGRs, I review the observational properties of
the twenty  AXPs\slash SGRs currently known  and describe the main features of the magnetar model.
In the last part of this review I discuss
the recent discovery  of magnetars with low external dipole field and some of
the  relations between  AXPs\slash SGRs and other classes of isolated  neutron stars.

\keywords{Neutron stars \and Magnetars }
\end{abstract}

\section{Introduction}
\label{intro}

The relevance of magnetic   energy in powering the emission from neutron stars (NSs), suggested more than four decades ago \citep{pac67},
has been well established with the discoveries  of anomalous X-ray pulsars (AXPs) and  soft gamma-ray repeaters (SGRs).
These are spinning-down, isolated NSs characterized  by the emission of
powerful bursts and flares  and with luminosities in the soft and hard X-rays
greater than their rotational energy loss.

Historically, AXPs and SGRs were  divided into  two distinct classes reflecting the way they were discovered,
but many observations indicate that there are no substantial differences between them.
Most AXPs were discovered as bright
pulsars in the soft X-ray range ($<$10 keV) and initially not distinguished  from the more numerous population
of X-ray binaries powered by accretion.
It was then pointed out that their narrow period distribution, long term spin-down,  soft X-ray spectrum and faint
optical counterparts were at variance with the properties of  pulsars in massive binaries  \citep{mer95}.
SGRs were instead discovered in the hard X-ray/soft $\gamma$-ray
range through the observation of bright and short bursts and classified as  a subclass of gamma-ray bursts
\citep{lar86,att87}, with the notable property of ``repeating''
from the same sky direction. When   the persistent X-ray counterparts of
SGRs were identified, it was found that they are pulsating sources very similar to the AXPs.

Even though alternative interpretations have been proposed,
the model involving magnetars, i.e. highly magnetized NSs, is the one that  most successfully
explains the properties of AXPs and SGRs  and is currently widely accepted.
According to the magnetar \linebreak model, the emission from these sources is ultimately powered by the energy stored in
their strong magnetic fields, which reach B$\sim10^{13}$--10$^{15}$ G in the magnetosphere,
and likely even higher values in the NS interior.
This sets them apart from most other NSs  which are powered by rotational energy, accretion, or residual heat.

In the last few years it has been  realized that the observational distinction
between magnetars and  other  NSs is not as sharp-cut as previously considered and not solely based on the magnetic field intensity.
This was already suggested by the existence of some radio pulsars  with inferred values of B overlapping
those of AXPs\slash SGRs, but which never showed bursts, flares or other signs of magnetically-powered emission.
The recent discovery of NSs that, despite their relatively low external dipole fields,
can certainly be classified as AXPs\slash SGRs  based on their outburst properties, indicates that the important
discriminant is most likely related to the strength and geometry of the magnetic field in the NS interior,
which unfortunately is less prone to direct estimates.

In the first part of this paper I briefly describe why it is believed that the energy sources which are seen to operate
in other NSs cannot power the emission of AXPs\slash SGRs\footnote{Some authors distinguish between these two designations,
calling SGRs   only    the sources which showed episodes of many repeated and intense bursts.
In this review I will not make such a distinction and I refer to the whole group of objects as AXPs\slash SGRs.}. The
two following sections are devoted to the observational properties of these sources (Section 3) and
the main features of the magnetar model (Section 4). The discovery and implications of magnetars with low dipole field
and possible relations of AXPs\slash SGRs with other classes of isolated  NSs are discussed in the
last two sections. For other reviews on these sources see \cite{woo06,kas07,mer08,tur13}.

\section{Why a different energy source ?}

\subsection{Rotation}

The rotational energy of a NS with spin period  P=2$\pi$/$\Omega$   is  E$_{rot}=I\Omega^2$/2=2$\times10^{46}I_{45}P^{-2}$  erg,
where I = I$_{45}10^{45}$ g cm$^2$ is the star moment of inertia\footnote{The moment of inertia of a NS with
mass $M$ and radius $R$ is in the range $\sim(0.5-1)\times10^{45}$ ($M/\msun$) ($R/10 ~km)^2$ g  cm$^2$
for most equations of state.}.
Soon after the discovery of pulsars it was found that their spin periods are   increasing, implying a
loss rate of rotational energy, $\dot{E}_{rot}$, sufficiently high to power not only the pulsed radio emission,
but also the bright radio/optical/X-ray nebulae observed around the most energetic pulsars.
The first striking example was provided by the pulsar PSR B0531+21, whose $\dot{E}_{rot}$=4.6$\times10^{38}$ erg s$^{-1}$   matches the
total energy output of the surrounding Crab nebula\footnote{Only one pulsars more energetic than the Crab   is currently known:
the 16 ms pulsar in the Large Magellanic Cloud PSR J0537--6960, which has $\dot{E}_{rot}$=5$\times10^{38}$ erg s$^{-1}$.}.
Most of the spin-down power is lost in a Poynting-dominated wind rather than in beamed photons from the pulsar.
Despite the majority of rotation-powered pulsars is observed at  radio wavelengths (over 2,000), their energy output in this
band is only a very small fraction of $\dot{E}_{rot}$.
The efficiency is higher in  the X-ray band, where non-thermal emission is observed in about one hundred pulsars, with
an average efficiency
$\sim$10$^{-3}$,
but with a large dispersion around this  value \citep{pos02,li08}.
At $\gamma$-ray energies the efficiency approaches 100\% for middle aged-pulsars
(at these high values it is important to  specify the   solid angle of the radiation beam
used in computing the efficiency).

AXPs\slash SGRs have period derivatives $\pdot$ larger than those of radio PSRs, but their long periods lead to
values of  $\dot{E}_{rot}\sim10^{32}-10^{34}$ erg s$^{-1}$, too small to power their
luminosity\footnote{Some AXPs\slash SGRs reach quiescent luminosity levels below their  $\dot{E}_{rot}$,
but during outburst they are as luminous as the persistent ones.}.
This is a robust conclusion,  even when  luminosity  uncertainties deriving from the poorly known distances of some AXPs\slash SGRs
are considered, and it  rules out the possibility of interpreting these sources as rotationally powered NSs.

Of course, the energy budget argument to rule out rotation-powered emission does not apply if one assumes that AXPs\slash SGRs are isolated white dwarfs.
Due to its higher moment of inertia, a rotating white dwarf has  a
spin-down power 10$^5$--10$^6$ times larger  than that of a NS for the same P and $\pdot$ values.
This idea was first proposed for \ee , a 7 s  X-ray pulsar at the center of the supernova remnant CTB 109 \citep{mor88}, which
attracted attention well before the recognition of the AXPs as a separate  class of objects \citep{mer95}.
Other authors proposed models for AXPs\slash SGRs based on white dwarfs \citep{uso93,mal12}, but these are challenged by the short spin periods
observed in some objects and by the deep upper limits on the optical counterpart of \linebreak \zeroq\ \citep{dur11}.

\subsection{Accretion}

Mass accretion is a well established process in X-ray binaries.
The first extra-solar X-ray source, Sco X-1, discovered
in the rocket experiment that marked the beginning of X-ray astronomy 50 years ago \citep{gia62},
is an accreting NS\footnote{Sco X-1 was the second observed NS,
the first one being the NS in the Crab nebula, shining  as a 16 magnitude object in the visible band.
Both were clearly understood as NSs only after the discovery of radio pulsars in 1968.}.
We now know several hundreds accretion-powered NSs in our Galaxy, as well as in other galaxies.
All of them are accreting  matter provided by their companion stars in binary systems.

Some of the X-ray sources now belonging to the AXP group were once  believed to be  powered by
accretion from a binary companion, the most natural explanation at that time, based on analogy with the other
X-ray pulsars.
However, deep observations in the optical and near infrared (NIR) failed to detect the luminous counterparts expected
if these pulsars were in high-mass binaries.
Furthermore, X-ray timing studies did not show the orbital Doppler shifts expected for binary motion,
ruling out also low-mass systems \citep{mer98,wil99}.
Optical/NIR counterparts  have now been firmly identified (on the basis of correlated variability),
or proposed  (on the basis of accurate positional coincidence) for several AXPs\slash SGRs.
In most cases they are sufficiently faint to rule out normal companion stars  and the presence of accretion disks.
In three sources (see Table 1) the optical emission is modulated at the period observed at X-rays, implying that
it comes from the same rotating object and not from a binary companion.

All these findings strongly indicate that AXPs\slash SGRs are isolated NSs.
They could  be powered by accretion only if matter is provided either by the interstellar \linebreak medium (ISM) or by a   disk of
fall-back material produced in the supernova explosion from which the NS originated.
The former possibility is ruled out because, for typical  NS velocities  and ISM densities, the resulting accretion rate is  too small.
Models invoking residual disks around isolated NSs are instead more plausible and represent the most widely discussed alternative to the magnetar
interpretation \citep{cha00a,ert09,tru13}.
In this class of models various mechanisms for the disk formation and different origins for
the observed X-ray luminosity have been considered.
The interaction with the fossil   disk is also invoked to account for the observed long periods and  rapid spin-down values.
According to \cite{alp01}, who proposed a scenario which unifies  the
different classes of isolated NSs,
the initial properties of a fall-back disk are among the fundamental parameters
that determine the fate of a NS.

The main objection to  models based on accretion onto isolated NSs is that they cannot easily account for the bursts and flares observed
in AXPs\slash SGRs, thus
requiring some additional mechanism and/or energy source  to explain these phenomena.
One  possibility, often considered in so-called ``hybrid models'', is that the strong magnetic
field powering  the bursts is not dipolar, but it is
only present in higher order multipoles dominating near the NS surface.
They differ from the magnetar model because accretion is invoked to explain the AXPs\slash SGRs persistent X-ray emission.

\subsection{Residual heat}

At birth, NSs have internal temperatures of $\sim$10$^{11}$ K at birth, that rapidly drop to $\sim$10$^{9}$ K.
The dominant cooling mechanism for the following $\sim$10$^5-10^6$ years is neutrino emission from the star's isothermal core.
This leads to surface temperatures of several $10^5$ to $10^6$ K, with emission peaking in the soft X-ray band (see, \textit{e.g.}, \citep{zav09}).
Temperature gradients on the star's surface  modulate the observed flux at
the rotation period. Thermal X-ray emission can be detected in isolated NS with ages of $\sim10^4-10^6$ years,
provided they are sufficiently close and the X-rays  not too absorbed by the ISM.
Older NSs are too cool to significantly emit in the X-rays band, while in the youngest pulsars the thermal radiation is difficult to detect
because it is outshined  by the brighter non-thermal emission powered by $\dot{E}_{rot}$.

There are two relevant observational properties of AXPs\slash SGRs
that cannot be explained resorting only to the NS thermal energy content: (i) the  production of a variety of strongly
variable and energetic events, ranging from short bursts to  giant flares flares,
and (ii) the presence of  hard tails  in the spectra of the persistent emission.
Short bursts of hard X-rays,   historically   the defining characteristics of SGRs, have now been observed
in virtually all the AXPs. Their durations ($\sim0.01-1$ s), hard spectra (characteristic temperatures $\sim$30--40 keV)
with little or no spectral evolution,  and high peak luminosity (up to $\sim10^{42}$  erg s$^{-1}$)
cannot be explained by the emission of the NS residual heat.
Persistent emission with power-law spectra extending up to hundreds of keV has been discovered with INTEGRAL in several AXPs\slash SGRs \citep{kui04,kui06,goe06b}.
The luminosity in these (pulsed) hard tails  is a significant fraction of the total energy output from AXPs\slash SGRs and requires the
presence of non-thermal phenomena in their magnetospheres.

\subsection{Magnetic energy}

As discussed above, the main powering mechanisms\footnote{Another energy source is provided by nuclear reactions.
This powers the type I bursts observed in many accreting low mass X-binaries. The properties of SGR bursts are very different
from those of type I bursts.}
at work in other kinds of NSs have problems to explain the  properties of AXPs\slash SGRs.
This motivated interest in the alternative proposal of a magnetically-powered emission and led
to the  development of models based on magnetars \citep{dun92,tho95,tho02}.

The magnetic energy of a  NS with field B$_{15}$=10$^{15}$ G filling its entire volume is 3$\times$10$^{47}B_{15}^2$ erg
(for a NS radius of 12 km).
This is enough to power a luminosity  of $\sim$10$^{35}$ erg s$^{-1}$  for $\sim$10$^5$ years,
a plausible age for AXPs\slash SGRs, considering their frequent association  with supernova remnants (SNRs)
or young clusters of massive stars\footnote{The giant flares (section 3.1) in which up to $\sim10^{46}$ ergs can be released,
are energetically more challenging. This limits the number of such events that a magnetar can emit in its lifetime.}.
Slightly higher fields and/or long-lasting episodes with low luminosity (as in transients) make the energy budget adequate for
all the magnetar candidates observed to date.

There are several pieces of  evidence that indicate the presence of a high magnetic field in AXPs\slash SGRs.
The conventional way to estimate the magnetic field of isolated NSs is based on the relation

\begin{equation}
B_d  =  3.2\times10^{19}\sqrt{P \pdot} ~~~ G
\end{equation}

\noindent
which gives the dipolar field as a function of the spin period and its time derivative\footnote{B$_d$ is the value
on the star surface at the magnetic equator. The surface field at the pole is a factor of two larger.}.
This relation is based on the simplified and unrealistic assumption that the observed spin-down is caused by
the emission of a rotating dipole in vacuum (with orthogonal magnetic and rotation axes).
More realistic models of the NS magnetosphere give values within a factor of two
from the previous expression  (\textit{e.g.,} \citep{spi06}).

Equation (1) yields  for the AXPs\slash SGRs  magnetic fields  higher than those of typical radio pulsars
and up to  B$_d$$\sim$10$^{15}$ G (see Table \ref{tab:prop}).
Although often quoted as the evidence that these objects are magnetars, this argument \textit{by itself} is rather weak and
subject to criticism. For example, other processes, such as the ejection of a wind of relativistic particles,
can contribute to the observed torque \citep{har99}.

Indeed, the evidence for  strong magnetic fields powering the AXPs\slash SGRs,  supported by Eq. (1),  comes from
several concurring observations, of which the most compelling are related to the
extreme properties of the giant flares of SGRs.
Giant flares start with a short, extremely bright spike, lasting a
fraction of a second, during which photons reaching MeV energies are emitted with a hard spectrum.
This is followed by a tail with a softer spectrum, lasting several minutes and pulsed at the NS rotation period.
The short duration of the initial spikes is consistent with the propagation time of the
magnetic instability over the whole NS surface with Alfv\'{e}n speed  \citep{tho95}.
The confinement of the hot plasma responsible for the pulsating tails requires the presence of a strong magnetic field,
and sets a lower limit of the order of a few 10$^{14}$ G on its intensity.
Although only three giant flares have been observed, the sources which showed such rare events are, for what concerns all
the other properties, undistinguishable from the other SGRs/AXPs.

Other motivations for a high magnetic field come from the short bursts, which
have now been detected from almost all  AXPs\slash SGRs. The peak luminosity
of the bursts exceed by a few orders of magnitude the Eddington limit for a NS. This is possible
because the intense magnetic field affects electron scattering, which becomes
strongly anisotropic and dependent on the photon polarization mode.
The cross section of the extraordinary mode (\textbf{$\bar{E}$} $\cdot$ \textbf{$\bar{B}$}=0) photons with angular
frequency $\omega$ is reduced by
a factor ($e B$/$\omega m_e c)^2$ with respect to the Thomson value.
As a consequence the radiative flux can exceed the classical Eddington limit in the presence of a strong magnetic field.

Another  evidence for very high magnetic fields in SGRs has been pointed out by \citep{vie07}, who
considered the high-frequency quasi periodic oscillations (QPOs) observed during the giant flare of
\zerosei . The QPOs at  625 and 1840 Hz involved
extremely large and rapid luminosity variations, with
$\Delta$L/$\Delta$t as large as several 10$^{43}$ erg s$^{-2}$.
This value exceeds the  luminosity-variability limit $\Delta$L/$\Delta$t $<$ $\eta$~2$\times10^{42}$  erg s$^{-2}$,
where $\eta$ is the efficiency of matter to radiation conversion
\citep{cav78}.  The relativistic effects, generally invoked to circumvent this limit (e.g. in blazars and gamma-ray bursts) are
unlikely to be at work in the SGR QPO phenomenon. \citep{vie07} instead propose that the Cavallo-Rees limit does not apply thanks
to the reduction in the photon scattering cross section induced by the strong magnetic field. In this way a lower limit of
$\sim2\times10^{15}$ G $(10~\textrm{km} / R_{NS})^3$ $(0.1/\eta)^{1/2}$ for the surface magnetic field is derived.

\begin{table*}
\caption{Multi-wavelength detections of confirmed AXPs and SGRs (P=pulsed,
D=detected, T=transient). The distances are in several
cases uncertain; the Table reports the values assumed in this
paper. MSC = massive stars cluster.}
\label{tab:sample}
 \begin{tabular}{lcccccccrl}
\hline
  &  X-rays    & X-rays     & Opt. & NIR & MIR & Radio & D       &   N$_{H}$    &  Location/association      \\
  &  $>$10 keV & $<$10 keV  &      &     &     &       &  (kpc)  & (cm$^{-2}$)  &    References        \\
  \hline
\oo      & D?& P   & P  & D  & -  & -  & 9   & 6 10$^{21}$   &                  \cite{ley08,mer95,dhi09,isr02,tie05a}  \\
         &   &     &    &    &    &    &     &               &         \\
 \hline
\qui     & P & P,T & -  & D  & -  & P  & 5   & 5 10$^{22}$   & G327.24--0.13      \\
         &   &     &    &    &    &    &     &               & \cite{gel07,hal08,cam07c,isr09,kui09,tie10} \\
 \hline
\kes     & P & P   & -  & D? & -  & -  & 8.5 & 2.3 10$^{22}$ & Kes 73           \\
& & & & & & & & &  \cite{vas97,kui04,tes08}\\
 \hline
\ee      & - & P   & -  & D  & D  & -  & 3.2 &  10$^{22}$   & CTB 109            \\
& & & & & & & & &  \cite{mer95,hul01,kap09,kot12} \\
 \hline
\rxs     & P & P   & -  & D? & -  & -  & 3.8 & 1.4 10$^{22}$ &                 \cite{tes08,sug97,den08b,rea05a} \\
& & & & & & & & &   \\
 \hline
\uu      & P & P   & P  & D  & D  & -  & 3.6 & 5 10$^{21}$   &                  \cite{mer95,den08a,hul04,ker02,wan06,rea07e}  \\
& & & & & & & & &   \\
 \hline
\smc     & - & P   & -  & -  & -  & -  & 61  & 6 10$^{20}$   & SMC              \\
& & & & & & & & &  \cite{lam02,tie08} \\
 \hline
\cxo     & - & P,T & -  & -  & -  & -  & 3.9 & 1.3 10$^{22}$ & MSC               \\
& & & & & & & & &  \cite{mun06,isr07b} \\
 \hline
\ctb     & - & P   & -  & -  & -  & -  & 13.2& 2.5 10$^{22}$ &  CTB 37B          \\
& & & & & & & & & \cite{hal10b,tia12} \\
 \hline
\psr     & - & D,T &  - & -  & -  & P  & 9   &   5 10$^{22}$ & G333.9+0.0        \\ 
& & & & & & & & & \cite{lev10,and12} \\
 \hline
\zeroq   & - & P,T & -  & -  & -  & -  & 2   & 1.1 10$^{21}$ &                  \cite{van10,esp10} \\
& & & & & & &  & &  \\
\hline
\zeroc   & P & P,T & P  & P  & -  & T  & 1.5 & 9 10$^{21}$   &                  \cite{gog08,rea09,apt09,dhi11,eno10} \\
& & & & & & & & &  \\
 \hline
\lmc     & - & P   & -  & -  & -  & -  & 55  & 5 10$^{21}$   & LMC, N49        \\
& & & & & & &  & &  \cite{tie09}  \\
 \hline
\sedici  & - & P,T & -  & -  & -  & -  & 11  & 9 10$^{22}$   &                  \cite{woo99c,esp09a,esp09b}  \\
& & & & & & &  & &  \\
 \hline
\zerosei & D & P   & -  & D  & -  & T  & 10  & 6 10$^{22}$   & MSC              \\
& & & & & & & & & \cite{kou98,mer05a,isr05} \\
 \hline
\diciotto& - & P,T & -  & -  & -  & -  & 10  &  10$^{23}$   &                  \cite{gog10,esp11} \\
& & & & & & &  & &  \\
 \hline
\zerozero& D & P   & -  & D? & -  & T  & 15  & 2 10$^{22}$   & MSC               \\ 
& & & & & & & & &  \cite{goe06b,hur99e,tes08} \\
 \hline
\swia    & - & P,T & -  & -  & -  & -  & 5   & 2 10$^{21}$   &                  \cite{liv11,rea12,sch12}      \\
& & & & & & &  & &  \\
 \hline
\swib    & - & P,T & -  & -  & -  & -  & 5   & 1.3 10$^{21}$ & W41                 \\
& & & & & & & & & \cite{kar12,esp13} \\
 \hline
\xte     & - & P,T & -  & D  & -  & P  & 3.1 & 6 10$^{21}$   &                 \cite{ibr04,isr04a,cam06,got07b}    \\
& & & & & & &  & &  \\
 \hline
\end{tabular}
%
\end{table*}


\section{Observational properties}

Twenty confirmed AXPs\slash SGRs are currently known (Table 1).
With the exception of \lmc\ and \smc , which are in the Magellanic Clouds, all of them are Galactic sources distributed at low latitudes in the
Galactic plane.
Although precise distances are generally not available, their  spatial distribution
points to typical distances of several kiloparsecs.
This is confirmed by the  associations with SNRs which have been proposed for several AXPs\slash SGRs. The most reliable ones, i.e.  those of the
AXPs\slash SGRs located at (or close to) the center of the remnant, are indicated in  Table~\ref{tab:sample}.
Three sources are likely located in clusters of massive stars (marked as MSC in  Table~\ref{tab:sample}),
suggesting that magnetars are formed in the collapse of very massive stars.  The associations
with SNRs and young star clusters, as well as their distribution
in the Galactic plane, indicate that AXPs\slash SGRs are  relatively young NSs.

AXPs\slash SGRs have spin periods in the range 2--12 s
and positive period derivatives between $10^{-12}$  and  $10^{-10}$   s s$^{-1}$ (Table~\ref{tab:prop}), with  few notable exceptions discussed below.
Their spin-down is not uniform: it is affected by timing noise
and large variations of $\pdot$  on short timescales have been observed in a few cases.
Many AXPs\slash SGRs, for which accurate timing measurements spanning months or years could be obtained,
showed spin-up glitches with values of $\Delta$P in the same range of those of rotation-powered pulsars,
despite their longer periods.
The glitches are sometimes associated to changes in the emitted radiation (e.g. bursts, variations in the flux or pulse profile),
but there is no simple one-to-one correlation between timing and radiative events.
Glitches apparently not associated with bursts and/or flux variations, and vice versa,  have been observed \citep{dib07a,zhu10,pon12}.
A spin-down glitch (i.e. $\Delta$P$>$0, contrary to the usual situation) might have occurred during the August 1998 giant flare
of \zerozero\ \citep{woo99b}, although it is also possible that the observed period change was due to a strong increase in the spin-down
before, or immediately after,  the flare rather than to a glitch.

Half of the AXPs\slash SGRs are persistent X-ray sources, i.e. they have always been detected at a nearly constant  X-ray luminosity
of $\sim$10$^{35}-10^{36}$ erg s$^{-1}$.
The other ones, most of which were discovered in the latest years during bright outbursts
generally associated to the emission of short bursts,  can be classified as transient sources.
When bright, their X-ray spectra, luminosity and pulse profile have the same properties seen in the persistent AXPs\slash SGRs.
The decays from the outburst can have different duration and shape and are generally characterized by a spectral softening \citep{rea11}.
In quiescence, luminosities as low as $\sim10^{31}$ erg s$^{-1}$ have been observed.  The outburst duty cycle of transient
AXPs\slash SGRs is still poorly known: multiple outbursts have been  observed  only in \sedici\ and \qui\  \citep{esp08,ber11}.

The X-ray spectra of AXPs\slash SGRs are rather soft below 10 keV  and, with only  few exceptions, strongly affected
by the interstellar absorption at low energy. They are generally fit with two-component models consisting of a blackbody
of temperature kT$_{BB}\sim$0.5 keV
plus a power-law (photon index $\Gamma\sim$2--4)  or another blackbody.
Many AXPs\slash SGRs have been detected also in the hard X-ray range, with power-law tails extending up to $\sim$100-200 keV.
These hard tails are pulsed,  often  time-variable, and contribute a significant fraction of the total luminosity.
The upper limits in the MeV region require the presence of  a spectral cut-off. No detections at higher energies have
been reported with $\gamma$-ray satellites (Fermi, AGILE) or ground based Cherenkov telescopes.

Although searches for optical counterparts are  difficult  due to the high absorption and crowded fields,
faint optical and/or NIR counterparts have been identified for several AXPs\slash SGRs.
They  have K band magnitudes of $\sim$20--22, which sometimes vary showing either a flux correlation or  anti-correlation with the X-rays.
Pulsations at the NS spin period have been observed in the optical emission of three sources.
In general the  optical/NIR fluxes lie above the extrapolation of the blackbody-like X-ray spectra.
It should be kept in mind that such fluxes are dependent on the assumed reddening and, therefore, subject to significant uncertainties.
A notable exception is provided by \zeroq\  which is located in the Galactic anticenter direction at a relatively high latitude compared
to the other AXPs\slash SGRs (b=+5$^{\circ}$).
Deep observations with the Hubble Space Telescope provided limits of $\sim$28.6 and 27.4 on the magnitudes of its possible counterparts
in the V and J band, respectively \citep{dur11}.  These rule out models
involving isolated white dwarfs with realistic temperatures.

Detections in the mid infrared (MIR) band have been obtained
for  \uu\ at 4.5 and 8 $\mu$m \citep{wan06} and for \ee\ at  4.5  $\mu$m   \citep{kap09} with the Spitzer Space Telescope.
They have been interpreted as evidence for  a residual disk of debris from the supernova,
irradiated and heated by the magnetar's X-ray flux,  but not contributing to  the X-ray emission by
accretion\footnote{See  \cite{ert07} for a different interpretation  in terms of an accretion-based model.}.

Two  different phenomena have  been observed in \linebreak AXPs\slash SGRs at radio wavelengths:  variable emission related to the ejection of
relativistic matter during  giant flares (see Section 3.1) and pulsed emission
in \linebreak  \xte , \qui\ and \psr .
The fact that these three sources  are transients  led to the speculation that the mechanisms responsible for the pulsed radio emission
in AXPs\slash SGRs might be related to their transient nature. However, no radio detections of all the other transient AXPs\slash SGRs have been obtained yet.
The   pulsed radio emission of AXPs\slash SGRs is quite different  from that of radio pulsars: they show strong variability  on daily timescales,
their spectra are very flat with $\alpha>$--0.5  (where $S_{\nu} \propto \nu^{\alpha}$),
and their average pulse profiles change  with time \citep{cam07a,cam07c,cam08}.
Such differences suggest that the radio emitting regions are more complex than the dipolar open field lines
along which the radio emission in normal pulsars is thought to originate.
The extremely precise positional measurements affordable through radio interferometry has made it possible to
derive proper motions: 13.5  mas yr$^{-1}$   for \xte\ \citep{hel07}  and 9 mas yr$^{-1}$ for \qui\ \citep{del12}.
For the distances given in Table \ref{tab:sample}, the implied transverse velocities are
of 190 km s$^{-1}$ and
230 km s$^{-1}$,  respectively, well in the range of those of radio pulsars.

\subsection{Bursts and flares}
\label{burst}

AXPs\slash SGRs are characterized by highly variable radiative phenomena spanning a wide range of time scales and luminosities:
from the short bursts of soft $\gamma$-rays, which led to the discovery of the SGRs,
to the extremely energetic giant flares, observed to date in only three sources,
 \lmc\   (March 5, 1979; \cite{maz79}), \linebreak
 \zerozero\  (August 27, 1998; \cite{hur99}),
and  \zerosei\ (December 27, 2004; \cite{pal05,mer05b}).

The typical short bursts have peak luminosity of $\sim$10$^{38}$--10$^{42}$ erg s$^{-1}$ and durations
in the range $\sim$0.01-1 s, with a lognormal distribution peaking at $\sim$0.1 s.
They usually  consist of single  (or a few) pulses with   fast rise time  and    slower decay.
The bursts can occur sporadically, separated by long time periods, or in groups  of tens or hundreds concentrated
in a few hours. The periods of high bursting activity are often associated to enhancements of the persisitent
emission (outbursts) of transient AXPs\slash SGRs,
but there are also cases in which only one or a few isolated bursts have been observed, both in transient and persistent sources.
The bursts  fluences  span the  range from a few 10$^{-10}$ to $\sim$10$^{-4}$ erg cm$^{-2}$, and follow a power law
distribution \citep{gog00,apt01,goe06}.
The burst spectra in the hard X-ray range (above $\sim$15 keV) are well fit by optically thin thermal bremsstrahlung models with
temperature  kT$\sim$30--40 keV.
When also lower energy data are included, a model consisting of the sum of two blackbody functions,
with temperatures kT$_1$$\sim$2-4 keV and kT$_2$$\sim$8-12 keV, provides a better description of the burst spectra \citep{fer04,oli04,esp07b,isr08}.

Giant flares involve the sudden release of an enormous amount of energy ($\sim$(2--200)$\times10^{44}$ ergs)
and have  unique spectral and timing signatures:  they start with a short hard spike followed by a longer pulsating tail.
Their properties indicate that a large fraction of the energy escapes directly as a relativistically expanding electron/positron plasma,
while the rest is gradually radiated away by a thermal fireball magnetically trapped in the NS magnetosphere which gives rise to the pulsating tail.
Other features that can be present in giant flares  are a precursor burst and long lasting afterglows at radio (see below)
and soft $\gamma$-rays  \citep{mer05b}.

The initial spikes of hard radiation reach a peak luminosity\footnote{Here and in the following we quote luminosities for isotropic emission.}
larger than 4$\times10^{44}$ erg s$^{-1}$ (up to $\sim$10$^{47}$ erg s$^{-1}$ for \zerosei ) and their
spectrum, with characteristic temperature of hundreds of keV, is much harder than that of short bursts.
The initial spikes  are characterized by  rise time smaller than a few milliseconds and  duration of a few tenths of second.
A complex, structured profile has been observed in the initial spike of the  2004 giant flare of \zerosei\ \citep{ter05,sch05}.

The pulsating tails of giant flares display a strong evolution of the flux, timing and spectral properties.
They have optically thin bremsstrahlung spectra with  temperatures of a few tens of keV, but
in the two most recent and better observed giant flares a combination of cooling thermal components and power laws extending into the MeV region was required
\citep{gui04,bog07,fre07}.
The decaying light curves, lasting several  minutes, are strongly modulated at the NS rotation period and show complex pulse profiles
which evolve with time.

Despite the energy emitted in the initial spike of \zerosei\ was larger  than that of  the other sources,
the energy  in the pulsating tails  was roughly of the same order ($\sim10^{44}$ ergs) for the three giant flares.
Since the tail emission originates from the fraction of the initial energy
that remains trapped in the NS magnetosphere, forming an optically thick photon-pair plasma
\citep{tho95}, this indicates that the magnetic field in the three sources is similar.
In fact the amount of energy that can be confined in this way is determined by the magnetic field strength,
which is thus inferred to be of several 10$^{14}$ G in these three magnetars.

Quasi periodic oscillations (QPOs) were discovered in the giant flare of  \zerosei\  \citep{isr05b}.
This prompted a re-analysis of the data available for the previous giant flares and the same phenomenon was found in \linebreak \zerozero\  \citep{str05}.
The QPOs  appear at  frequencies in the range $\sim$20-150 Hz
(and also at 625 and 1840 Hz for \zerosei )   for different time intervals during the tails of the giant flares,
and are often  correlated with specific phases of the spin-period.
They have been attributed to seismic vibrations in the NS  solid crust or to Alfv\'{e}n oscillations in the fluid
core (see, \textit{e.g.}, \cite{wat11} and references therein). Their study could provide constraints on the NS internal structure.

Transient radio emission was detected after the  giant flares of \zerozero\  and \zerosei\ \citep{fra99,gae05}.
The properties of the \zerosei\ radio emission could be studied in detail for more than one year. The proper motion of the radio blob
indicates an anisotropic ejection of relativistic matter
in a solid angle of $\sim$0.5--1 sterad.
Other parameters could be estimated by modelling the source expansion and flux temporal evolution.
Although  jet models with  different  physical and geometrical details can fit the data equally well, all of them show
evidence for an anisotropic ejection of 10$^{24}$--10$^{25}$ g of mildly relativistic matter associated to the giant flare  \citep{cam05,gra06}.

A few strong flares, involving less energy than the giant flares, but definitely brighter and much rarer than the normal
short bursts, have also  been seen in a few sources.
The strongest of these ``intermediate'' flares  was observed on April 18, 2001 from \zerozero\ \citep{kou01,gui04}.
It  lasted about 40 s and showed pulsations at the NS rotation period, as in the tails of giant flares, but it lacked an initial spike.
Particularly strong bursts
have also been observed, for example  in \sedici\ \citep{maz99c}  and  in \qui\ \citep{mer09}.
It is possible that all these bursts and/or flares can be explained by the same physical process, involving
a large distribution of energies \citep{mer11b}. It seems, however, that the giant flares, especially due to the peculiar properties
of their initial spikes, are a different phenomenon.

\begin{table*}
\caption{Main properties of the AXPs and SGRs. Flux is in units of 10$^{-11}$ erg cm$^{-2}$ s$^{-1}$ and is not corrected for the absorption.
Luminosity corrected for the absorption in units of 10$^{35}$ erg  s$^{-1}$.  F=Giant Flare, B=burst(s),
G=glitch(es).}
\label{tab:prop}
\begin{tabular}{lccccccl}
 \hline
 &Period& $\pdot$      & B        &  Flux                     &  L$_{X}$  &  Notes &  References     \\
 &  (s)  &  (10$^{-11}$ s s$^{-1}$)  &  (10$^{14}$ G) &   [2--10 keV] & [0.5--10 keV]           &   &       \\
  \hline
\oo    & 6.45 & 1--10 &2.6--8.1 & 0.5--4  & 0.9--7              & B,G   & \citep{mer95b,tie05a,gav02,dib09}  \\ 
\qui   & 2.07 & 2.3   & 2.2     & 0.03--9 & $\sim$0.01--4       & B,G   & \citep{esp08b,mer09,ber11,kui12}   \\ 
\kes   & 11.77& 4.1   & 7       & 1.6     &  4                  & B,G   & \citep{got02,mor03,dib07a,kum10b,lin11}  \\
\ee    & 6.98 & 0.048 & 0.6     & 1--4    & 2--7                & B,G   & \citep{woo04,gav04,kas03} \\ 
\rxs   & 11.0 & 2.4   & 5.2     & 2--3    & 1.5--2.5            & G     & \citep{rea05a,isr07,dal03,dib07a}  \\ 
\uu    & 8.69 & 0.2   & 1.3     & 6--7    & 3                   & B,G?  & \citep{isr94,gav08,rea07e,gon10,gav11} \\ 
\smc   & 8.02 & 1.9   & 4       & 0.013   & 2                   & -     & \citep{lam02,mcg05,tie08}\\  
\cxo   & 10.6 &$<$0.04 & $<$0.7 & 0.01--4 & 0.007--$\sim$1      & B,G?  & \citep{mun07,woo11,an13} \\ 
\ctb   & 3.82 & 6.4   &  5      &  0.14   & $\sim$1             & -     & \cite{hal10b,hal10c,sat10} \\ 
\psr   & 4.33 & 1.7   & 2.7  & 0.003--0.14 & $\sim$0.01--40     & -     & \citep{and12}  \\ 
\zeroq & 9.1 & 0.0004 & 0.06    & 0.002---3 & $\sim10^{-4}$--20 & B     & \citep{rea13,esp10} \\
\zeroc &  5.7 & 0.7   & 2       & 0.13--4 & $\sim$0.01--0.25    & B     & \citep{rea09,eno09} \\ 
\lmc   & 8.05 & 4     & 5.7     & 0.05    & $\sim$10            & F,B   & \citep{maz79,maz99b,tie09} \\ 
\sedici & 2.59 & 1.9   & 2.2   & 0.006--1.2& $\sim$0.01--2      & B     & \citep{mer06a,esp09a,esp09b} \\ 
\zerosei & 7.6 & 8--80 & 8--25   & 1--3    & 2--6               & F,B   & \citep{pal05,mer05c,woo07}\\ 
\diciotto& 7.56& 0.28  & 1.5 &  $<$0.004--0.4 &   $<$0.01--1    & B     & \citep{gog10,esp11} \\ 
\zerozero& 5.2 & 5--14 & 5.1--8.6& 0.4--1  & 2--5               & F,B,G? & \citep{maz79b,woo99b,mer06b} \\ 
\swia   & 8.44 & 0.03 & 0.5  &  0.008--25 & 0.002--10           & B     & \citep{rea12,sch12} \\  
\swib   & 2.48 & 0.8  &  1.4 &  $<$0.0004--3 &  $<10^{-4}$--1   & B     & \citep{kar12,esp13} \\ 
\xte    & 5.54&0.8--2.2& 2.1--3.5& 0.05--8 & 0.008--1.3         & B    & \citep{got07b,woo05,ber11b}  \\ 
\hline
\end{tabular}
\end{table*}

\section{The magnetar model}

Strongly magnetized NSs
can form if they are born with initial spin period $P_{\circ}$ shorter than the overturn time of $\sim$3--10 ms of the
convection driven by the high neutrino luminosity L$_{\nu}>10^{52}$ erg s$^{-1}$ \citep{dun92}.
The efficient dynamo resulting in this case can generate magnetic fields as high as $3\times10^{17} (P_{\circ}/1~ms)^{-1}$ G.
Rapid neutrino cooling in the proto-NS is essential in driving the strong turbulent convection which amplifies the seed field.
Such a dynamo operates only for a few seconds, but, in principle, it can generate fields  as high as 10$^{16}$ G,
most likely with a multipolar structure and with a  strong toroidal component, in the NS interior \citep{tho02,cio10}.

An alternative formation scenario has also been proposed, according to which
magnetars would be the descendent of the stars with the highest magnetic fields.
The wide distribution of field intensities in magnetic white dwarfs has been attributed to
the spread in the magnetic fields of their progenitors and a similar situation  could apply to NSs \citep{fer06,hu09}.
The possible  association of some AXPs\slash SGRs with
clusters of massive stars seems to indicate that they descend from stars of mass above $\sim$40 $\msun$.
However, evidence for a lower mass progenitor has been derived for  \zerozero\  \citep{dav09}.

Independent on the origin of their strong field, young magnetars undergo rapid spin-down due to magnetic dipole radiation and particle wind losses.
Rotation periods of several seconds are reached in a few thousands years, explaining why no magnetars are observed at short periods.

Magnetic field decay provides a  source of internal heating which can play an important role in the
X-ray emission from the surface of magnetars.
This internal heating source yields a surface temperature higher than
that of a cooling NS of the same age and smaller magnetic field. The enhanced thermal conductivity in
the strongly magnetized envelope  contributes to raise the surface temperature \citep{hey97a}.
The decay of the magnetic field is in turn affected by the NS temperature evolution, which makes it difficult to
derive a self-consistent, complete and realistic model. Simulations in 2-D have been performed for the case
in which the magnetic field is  sustained by currents in the NS crust \citep{pon09}.
They indicate that systematically higher surface temperatures are obtained for NSs with strong internal
toroidal fields.

The evolution of the internal field deforms  the NS crust contributing to the persistent X-ray emission \linebreak through low
level seismic activity and storing magneto\slash elastic energy which becomes available to
power bursts and flares when the crust cracks.
Alfv\'{e}n waves in the magnetosphere   accelerate charged particles leading to Comptonization and
bombardment of the surface   \citep{tho96}.
A first attempt to derive the overall bursting properties of magnetars, taking into
account the magneto/thermal evolution of the poloidal and toroidal internal field components, indicates
significant differences in the energetics and recurrence time as the magnetar evolves \citep{per11}.
Objects with a lower toroidal field are generally less active.
The frequency of the bursts  decreases with age and, to a lesser extent, also their energy. Interestingly, sporadic
bursts are predicted for magnetic fields as low as a few 10$^{12}$ G.

A twisted  magnetosphere ($B_\phi\neq 0$) supports currents much larger than the Goldreich-Julian current flowing along
open field lines in normal pulsars \citep{tho00,tho02}.
An electron/positron corona is formed in the closed magnetosphere,   consisting of flux tubes  anchored on both ends to the NS surface.
The currents are carried by charges extracted from the NS surface and pairs produced along the flux tubes.
The strong flow of charged particles  provides an additional source of heating for the star surface and
gives rise to a significant optical depth for resonant cyclotron scattering in the magnetosphere.
Repeated scattering of the thermal photons emitted at the star surface can produce the  power-law components observed
in the spectra of AXPs\slash SGRs  \citep{lyu06b,fer07,nob08a}.

Most of the models developed to derive the spectral and timing evolution properties of magnetars
considered globally twisted magnetospheres, but it is more likely that the twist is imparted only  to a small
bundle of field lines.
How the magnetic energy is dissipated in an  untwisting  magnetosphere has been studied in detail by \cite{bel09}, who derived
expressions for the resulting luminosity evolution. Most of the energy is dissipated as thermal radiation in the
footprints of a bundle of twisted lines which are heated by the accelerated magnetospheric particles. The magnetosphere
gradually untwists through the expansion of a potential region which causes a reduction in the area of these hot
spots. The small emitting surfaces inferred from the blackbody  fits of transient AXPs\slash SGRs
are consistent with this scenario.

An alternative possibility  to explain the transient magnetars involves the fast deposition of energy  deep in the NS crust \citep{lyu02}.
This could result, for example, from  a sudden fracture driven by magnetic stresses and would induce a heating
of the surface.
Since the subsequent thermal evolution depends primarily on the properties of the outer crust and on
the depth of the energy deposition, a  modelling of the outburst decays might provide some information on the star structure.
However, the transients observed up to now are far from showing a uniform picture \citep{rea11}. Furthermore,
the systematic and statistical spectral uncertainties affecting the observations
often prevent a simple comparison with the predictions of the models, as shown, \textit{e.g.},
by the case of \sedici\ \citep{mer06a}. Recent theoretical work suggests that the  maximum luminosity
observable in an outburst is  limited to $\sim$10$^{36}$ erg s$^{-1}$, independent on the amount of energy released in the
crust  \cite{pon12}. This is due to the self-regulating effect resulting from the strong temperature-dependence of neutrino emission
and  might explain why persistently bright AXPs\slash SGRs do not undergo outbursts, but change their luminosity by, at most, of a factor of a few.

It is clear that the giant flares must be powered by magnetic energy, but how this exactly occurs  and where the
energy is accumulated before being released remain open questions. The two main scenarios which have been considered
involve either the build-up of elastic energy in the crust, which finally causes a large-scale fracture \citep{tho95,tho01}, or
the gradual injection of energy in the magnetosphere, which is released when an instability produces a rearrangement of the field
\citep{lyu06a,gil10,yu12}. In the first class of models the flare occurs when the tensile strength of the crust is exceed,
while in magnetospheric models it is more difficult to determine which is the triggering process that can explain the extremely
rapid energy release. More data will certainly help to answer these questions, but the extremely rare and completely
unpredictable nature of these events limit the observational progress.

\section{Magnetars with low external field}

Until recently the known AXP/SGR with the lowest magnetic field (as given by Eq.\ 1) was \ee\ with B$_d$=6$\times$10$^{13}$  G,
only slightly larger than the quantum critical field\footnote{$B_{QED}$ is the magnetic field for which the energy of the first Landau
level of the electron equals its rest mass. It was often regarded as the boundary  between normal pulsars and magnetars, although
there is no real physical reason or threshold effect to motivate this.}
$B_{QED}=\frac{m^2 c^3}{\hbar e}$=4.4$\times10^{13}$ G.
This pulsar, despite  its value of B$_d$ smaller than that of all the other magnetar
candidates,  was the first AXP to show the strong bursting activity\footnote{Two
isolated bursts were detected in October and November 2001 from \oo\ \citep{gav02},
but \ee\ in June 2002 emitted more than 80 bursts in 4 hours, associated to an increase of
the persistent flux and   a glitch.} previously believed to be a prerogative of SGRs \citep{kas03}.

In June 2009 the new transient \zeroq\ was discovered through the detection of two short bursts seen by different satellites \citep{van10,esp10}.
Pulsations at 9.1 s were soon found in its X-ray emission, but despite an extensive monitoring of the outburst decay
in the following months, no significant $\pdot$ could be measured. This was quite unexpected because the fast spin-down rates of all
the other transient AXPs\slash SGRs were always evident after only a few days of observations. The upper limit of $\pdot$$<$6$\times$10$^{-15}$ s s$^{-1}$
obtained after two years implied B$_d$$<$7.5$\times$10$^{12}$  G \citep{rea10}, an unprecedented value for a source showing all
the typical characteristics of a magnetar.
Only recently, thanks to a phase-coherent timing analysis of all the observations of \zeroq\ spanning  more than three years,
it has been possible to measure its small spin-down of (4$\pm$1)$\times$10$^{-15}$ s s$^{-1}$ \citep{rea13},
which translates into a field   B$_d$=6$\times$10$^{12}$  G.

A similar situation occurred for another transient, \swia\ \citep{liv11}, discovered in July 2011 through the detection of several bursts,
while a timing analysis of more than six years of data  of \cxo\ led to revise its previously reported $\pdot$  into an upper limit \citep{an13}.
The dipole moments inferred for these two sources imply surface fields of B$_d$=5$\times$10$^{13}$  G
and   B$_d<$7$\times$10$^{13}$  G, respectively.
They are not as low as that of \linebreak \zeroq , but still below that of the bulk of ``traditional'' AXPs\slash SGRs and in the range
of fields inferred for a non-negligible number  of rotation-powered radio pulsars.

These findings indicate that a high magnetic dipole moment is not a mandatory condition for a magnetar.
Indeed what really matters to power the  magnetar  activity and emission is the strength of the internal  field, and in particular of its
toroidal component which is responsible for the NS crust deformation/cracking and for imparting twists to the
external magnetosphere. It is also likely that the magnetosphere is not dipolar but contains a significant
contribution from higher order multipoles, which increase the field close to the star surface, but fall-off rapidly
with radius and do not contribute to the spin-down.

It has been suggested that objects like \linebreak \zeroq\ are old magnetars,  in which a substantial decay of the magnetic field has occurred.
An initial configuration with internal toroidal field of a few 10$^{16}$ G  and external dipole of B$_d$=2.5$\times$10$^{14}$  G
can reproduce the observed properties of \zeroq\ in the context of a magneto-thermal evolution with field decay \citep{tur11}.
A weaker initial B$_d$ would not  spin-down the pulsar to the currently observed long period\footnote{Unless interaction
with  a residual disk is invoked  \citep{alp11}.}.
An immediate consequence of the external field decay is that the characteristic age, defined as $\tau_c$=P/2$\pdot$, greatly
overestimates the true age of the magnetar, which in the above model is of the order of $\sim$10$^6$ years.

\section{Is there evidence for magnetic activity in other classes of NSs ?}

As shown by  the sources discussed in the previous section, a high dipole field
is not a necessary condition for the onset of magnetar activity.
This leads to the interesting possibility that NSs with ``normal'' magnetic field values, as estimated with Eq. (1),  might show signatures
of magnetar-like behavior, such as, e.g., bursts or other kinds of variability.
Indeed some examples of this have been found in recent years, pointing to a closer connection and possible evolutionary links
among different classes of isolated NSs.

The 0.3 s pulsar PSR J1846--0258, at the center of the  Kes 75  SNR, was considered
a normal rotation-powered X-ray pulsar\footnote{Despite the lack of a radio detection, which could be due to an unfavorable orientation
of the radio beam.},
given its $\dot{E}_{rot}$ = 8$\times$10$^{36}$ erg s$^{-1}$   sufficiently high  to
power the  pulsar and nebula X-ray  emission (L$_X$ of 2.6$\times$10$^{34}$  and 1.4$\times$10$^{35}$ erg s$^{-1}$ , respectively).
In May 2008,  PSR J1846--0258 emitted four short bursts similar to those observed in AXPs\slash SGRs \citep{gav08b}.
The bursting activity was accompanied by a large spin-up glitch and marked the start of an enhancement
in the pulsed X-ray flux which lasted about one  month \citep{liv10}.
Spectral and flux variations associated with this event  were seen also
in the hard X-ray range \citep{kui09a}. All these phenomena are typical of  magnetars.
The dipole field inferred from the timing parameters of PSR J1846--0258 is B$_d$=5$\times$10$^{13}$ G.
It will be interesting to see if similar events occur also in other \textsl{bona fide} rotation-powered radio pulsars.
About one fifth of the radio pulsars with measured $\pdot$   have B$_d$ values larger than that of \zeroq .
This population could well harbor other ``hidden'' magnetars, possibly showing weak and sporadic bursting activity
(hence difficult to detect).

Magnetically-driven activity can manifest itself also through plastic deformations
of the  crust which  appear as long term variations in the (thermal) emission from the star surface, rather than as abrupt bursts.
A possible example has been observed in the 8.4 s X-ray pulsar \xdzerosette , which showed spectral
variability on a time scale of few years \citep{dev04,hoh09}.
Most of the variation occurred in a relatively short timescale of half a year, in coincidence with a possible glitch \citep{van07}.

This source belongs to the small group of X-ray Dim Isolated NSs
(XDINS\footnote{Also known with the ``Magnificent Seven'' nickname.})
discovered with the ROSAT satellite (see, \textit{e.g.}, \cite{tur09} for a review).
XDINS have very soft thermal spectra (blackbody temperatures T$_{BB}\sim$40--110 eV), X--ray luminosities
L$_X$$\sim$10$^{30}-10^{32}$ erg s$^{-1}$, spin periods in the 3--12 s range,
faint optical counterparts (V$>$25), and no detectable radio emission.
Their magnetic fields, inferred either from Eq. (1) for the few XDINS with measured $\pdot$ or from
the  absorption lines present in their spectra, are of the order of 10$^{13}$--10$^{14}$ G.
It is possible that the XDINS are old magnetars ($\sim$10$^5$--10$^6$ years),
in which most of the magnetic energy has been dissipated and whose X-ray emission is now powered by the residual thermal energy.
Their temperature  is consistent with theoretical cooling curves if the
the effects of stronger initial   fields  are taken into account \citep{agu08,pon07,pon11}.

Another example of long term variation which might be related to magnetic activity, is provided by the 0.4~s X-ray pulsar
RX J0822--4300, located in the Puppis A SNR.  XMM-Newton observations showed that the centroid
energy of an emission line visible in its X-ray spectrum decreased  from  0.8 keV in 2001 to 0.73 keV in 2009-2010 \citep{del12a}.
RX J0822--4300  belongs to a small group of sources known as Central Compact Objects (CCOs).
These are  X--ray sources  with thermal-like X-ray emission ( kT$_{BB}$$\sim$0.2--0.5 keV) located at the center of shell-like
SNRs and with  high ratios of X-ray-to-optical flux, typical of isolated NSs.
CCOs are not detected in the radio band, are not surrounded by pulsar wind nebulae and do not show any
evidence for  non-thermal components in their X-ray spectra (see,\textit{ e.g.}, \cite{del08}, for a review).
The CCOs properties  suggested an interpretation in terms of isolated
NSs with small values of $\dot{E}_{rot}$, despite the young age inferred from their associated SNRs.
This has  been  confirmed for
three CCOs in  which pulsations have been detected.  They have  periods in the 0.1--0.4 s range,
and exceptionally small spin-down rates  \citep{hal10,got13}, yielding
B$_d$=3.1$\times$10$^{10}$ G,  B$_d$=9.8$\times$10$^{10}$ G
and  B$_d$=2.9$\times$10$^{10}$ G,   for the sources in Kes 79, G296.5+10.0 and Puppis A, respectively.
These NS were probably born with spin periods very close to the current values and
it is likely that their relatively long initial spin period and low magnetic field be causally connected.
Alternatively, CCOs might have been born with a higher field which was then buried beneath the NS surface by the
accretion of fall back material after the supernova explosion. Depending on the amount of accreted
matter and on the properties of the NS crust and core, the   reemergence of the field
can take place on a large range of timescales, from a few thousand to several billion years \citep{ho11,vig12}.
The presence of a strong buried field can help to explain the gradients in the surface temperature
responsible for the large pulsed fractions and/or phase-dependent spectral properties of CCOs,
which are otherwise difficult to understand for their low values of B$_d$.
The variability in the X-ray spectral feature of RX J0822--4300, if related to magnetic activity,
supports this connection between CCOs and  AXPs\slash SGRs.

Finally, it is worth mentioning the Swift/BAT detection of two short SGR-like bursts  from the direction
of the peculiar X-ray/radio/$\gamma$-ray source LS I +61$^{\circ}$ 303 \citep{tor12,GCN12914}.
This is one of the very few X-ray binaries detected at GeV/TeV energies. It is composed of a Be type star and a compact object,
which is likely a NS (although this cannot be confirmed since no  pulsations have been detected and the possibility
of a black hole companion has also been considered).

\section{Conclusions}

The success of the magnetar model in explaining the properties of AXPs\slash SGRs indicates the existence of NSs
with magnetic fields of   $\sim$10$^{15}$ G.
Their extremely energetic flares yield the highest radiation fluxes which reach the Earth from outside
the Solar System, often causing measurable effects on the atmosphere \citep{ina99,tan10,nic12}.
Exciting  results have been obtained in the study of  AXPs\slash SGRs in the latest years, contributing to
change our vision of the population of NSs in the Galaxy.

Magnetars  are relevant also in other astrophysical contexts which have not been discussed here. These
include, for example,
the    supernova  explosions,
the production of gamma-ray bursts, the acceleration ultra-high-energy cosmic rays, the production of neutrinos,
and the emission of gravitational waves
(see, \textit{e.g.}, \citep{pir11,buc12,aro03,iok05,ste05}).
Being the only places in the Universe in which we can observe and study physical processes in  magnetic fields
of such extreme intensity,  AXPs\slash SGRs  will certainly remain among the most interesting astrophysical objects for theoretical
studies and for observations with the experimental facilities that will become available in the coming years.

\begin{acknowledgements}
I thank Paolo Esposito and Roberto Turolla for their careful reading
of this work and useful comments.
\end{acknowledgements}

\begin{small}

\bibliographystyle{plain}%

\end{small}

\end{document}